\documentclass[conference]{IEEEtran}
\IEEEoverridecommandlockouts

\usepackage[draft]{hyperref}
\usepackage{cite}
\usepackage{microtype}
\usepackage{graphicx}
\usepackage{subfigure}
\usepackage{lstautogobble}  
\usepackage{zi4}     
\usepackage{listings}
\usepackage{booktabs} 
\usepackage[linesnumbered,ruled,vlined]{algorithm2e}
\usepackage{soul}
\usepackage[skip=0pt]{caption}
\usepackage{rotating}
\usepackage{xspace}
\usepackage{color, colortbl}
\usepackage{xcolor}
\usepackage{minted}
\usepackage{multirow}
\usepackage[many]{tcolorbox}
\usepackage[T1]{fontenc}
\usepackage[scaled=0.92]{helvet} 
\usepackage{microtype}
\usepackage{xcolor}
\usepackage[table]{xcolor}
\usepackage{tcolorbox}
\usepackage{adjustbox}
\usepackage{makecell}
\usepackage{float}
\usepackage{listings}
\usepackage{xspace}
\usepackage{graphicx}
\usepackage{subcaption}
\usepackage{paralist}
\usepackage{microtype}
\usepackage{graphicx}
\usepackage{subfigure}
\usepackage{booktabs}
\usepackage{bigstrut}
\usepackage{enumitem} 
\usepackage{amsmath}
\usepackage{amssymb}
\usepackage{hyperref}
\usepackage{cleveref}
\usepackage{threeparttable}
\usepackage{ragged2e}

\definecolor{gray05}{gray}{0.95}
\definecolor{gray08}{gray}{0.92}
\definecolor{gray10}{gray}{0.90}
\definecolor{gray12}{gray}{0.88}
\definecolor{gray15}{gray}{0.85}
\definecolor{gray18}{gray}{0.82}
\definecolor{gray20}{gray}{0.80}
\definecolor{gray25}{gray}{0.75}
\definecolor{gray30}{gray}{0.70}
\definecolor{gray35}{gray}{0.65}
\definecolor{gray40}{gray}{0.60}
\definecolor{gray45}{gray}{0.55}
\definecolor{gray50}{gray}{0.50}
\definecolor{gray55}{gray}{0.45}
\definecolor{gray60}{gray}{0.40}
\definecolor{gray65}{gray}{0.35}
\definecolor{gray70}{gray}{0.30}
\definecolor{gray75}{gray}{0.25}
\definecolor{gray80}{gray}{0.20}
\definecolor{gray85}{gray}{0.15}
\definecolor{gray90}{gray}{0.10}
\definecolor{gray95}{gray}{0.05}
\definecolor{blue}{HTML}{0f62fe}
\definecolor{red}{RGB}{234,51,35}
\definecolor{ibm-blue-light}{HTML}{a6c8ff}
\definecolor{peach}{HTML}{ffdda6}
\definecolor{amber}{HTML}{cc4e00}

\definecolor{problemblue}{RGB}{100,134,158}
\definecolor{idiomsgreen}{RGB}{0,162,0}
\definecolor{exercisebgblue}{rgb}{0,  .69,  .941}
\definecolor{deepgreen}{rgb}{0.0, 0.5, 0.0}
\definecolor{codegreen}{rgb}{0,0.6,0}
\definecolor{codegray}{rgb}{0.5,0.5,0.5}
\definecolor{codepurple}{rgb}{0.58,0,0.82}
\definecolor{backcolour}{rgb}{0.95,0.95,0.92}
\definecolor{redColor}{RGB}{255,0,0}
\definecolor{Gray}{gray}{0.1}
\definecolor{javared}{rgb}{0.6,0,0} 
\definecolor{javagreen}{rgb}{0.25,0.5,0.35} 
\definecolor{javapurple}{rgb}{0.5,0,0.35} 
\definecolor{javadocblue}{rgb}{0.25,0.35,0.75} 
\definecolor{ibmblue}{RGB}{63,97,246}
\definecolor{maroon}{RGB}{105,33,61}
\definecolor{teal}{RGB}{59,115,115}
\definecolor{exercisebgblue}{rgb}{0,  .69,  .941}
\definecolor{bluekeywords}{rgb}{0.13, 0.13, 1}
\definecolor{greencomments}{rgb}{0, 0.5, 0}
\definecolor{redstrings}{rgb}{0.9, 0, 0}
\definecolor{graynumbers}{rgb}{0.5, 0.5, 0.5}
\definecolor{mygreen}{rgb}{0,0.6,0}
\definecolor{mygray}{rgb}{0.5,0.5,0.5}
\definecolor{mymauve}{rgb}{0.58,0,0.82}

\setitemize{noitemsep,topsep=0pt,parsep=0pt,partopsep=0pt,leftmargin=*, wide=0pt}
\setenumerate{noitemsep,topsep=0pt,parsep=0pt,partopsep=0pt,leftmargin=*, wide=0pt}
\usepackage{hyperref}

\definecolor{revblue}{RGB}{0,0,255} 
\newcommand{\REV}[1]{{\color{black}#1}}

\newcommand{\agent}{\textsc{Praxis}\xspace}
\newcommand{\graph}{PDG\xspace}
\newcommand{\graphs}{PDGs\xspace}
\newcommand{\sgraph}{SDG\xspace}

\newcommand{\tion}[1]{\S\ref{sec:#1}\xspace}

\newcommand{\be}{\begin{enumerate}}
\newcommand{\ee}{\end{enumerate}}
\newcommand{\bi}{\begin{itemize}}
\newcommand{\ei}{\end{itemize}}
\newcommand{\eg}{e.g.}
\newcommand{\ie}{i.e.}

\newcommand{\tab}[1]{\Cref{tab:#1}\xspace}

\lstdefinestyle{LogStyle}{
    basicstyle=\ttfamily\scriptsize,  
    keywordstyle=\bfseries,           
    columns=fullflexible,
    breaklines=true,
    backgroundcolor=\color{gray!10},
    frame=single,
    xleftmargin=15pt,
    framexleftmargin=15pt,
    showstringspaces=false,
    aboveskip=5pt,
    belowskip=5pt,
    literate={-}{-}1,                 
}

\lstdefinestyle{LogStyle}{
    basicstyle=\ttfamily\small,                
    keywordstyle=\bfseries\color{blue!70},     
    columns=fullflexible,
    breaklines=true,
    backgroundcolor=\color{gray!5},            
    frame=single,
    rulecolor=\color{gray!50},                 
    xleftmargin=20pt,
    framexleftmargin=20pt,
    showstringspaces=false,
    aboveskip=8pt,                             
    belowskip=8pt,
    frameround=fttt,                           
    framerule=0.5pt,                           
    literate={-}{-}1,                          
    escapeinside={(*}{*)}                      
}

\usepackage{boldline}

\setminted{
    frame=lines,
    framesep=1mm,
    fontsize=\footnotesize,     
    breaklines=true,
    breakanywhere=true,
    linenos=false,
    numbersep=0pt,
    escapeinside=||,
    mathescape=true,
    bgcolor=white,
    highlightcolor=magenta!12   
}

\usepackage{xparse} 

\NewDocumentEnvironment{tightminted}{O{}m} 
  {\vspace{-0.75\baselineskip}%
   \begin{minted}[#1]{#2}}
  {\end{minted}%
   \vspace{-0.5\baselineskip}}

\usepackage{amsthm}

\newcommand\mathllm[1]{$\mathcal{M}_{#1}$\xspace}
\newcommand\mathpdg{$G_{P}$\xspace}

\usepackage{cite}
\usepackage{amsmath,amssymb,amsfonts}
\usepackage{algorithmic}
\usepackage{graphicx}
\usepackage{textcomp}
\usepackage{xcolor}
\def\BibTeX{{\rm B\kern-.05em{\sc i\kern-.025em b}\kern-.08em
    T\kern-.1667em\lower.7ex\hbox{E}\kern-.125emX}}

\usepackage{booktabs,tabularx,array}
\newcolumntype{Y}{>{\raggedright\arraybackslash}X} 

\IEEEoverridecommandlockouts

\begin{document}

\IEEEpubid{\makebox[\columnwidth]{\parbox{\columnwidth}{
\vspace{50pt}
\rule{\columnwidth}{0.1pt} 
\footnotesize \copyright~2026 IEEE. Personal use of this material is permitted.  Permission from IEEE must be obtained for all other uses, in any current or future media, including reprinting/republishing this material for advertising or promotional purposes, creating new collective works, for resale or redistribution to servers or lists, or reuse of any copyrighted component of this work in other works.
\vspace{0.3em}
Accepted to appear in The 56th Annual IEEE/IFIP International Conference on Dependable Systems and Networks.
}
} \hspace{\columnsep}\makebox[\columnwidth]{ }}

\title{
\agent: Integrating Program Analysis with Observability for Root-Cause Analysis
}
\author{
\IEEEauthorblockN{Shengkun Cui\IEEEauthorrefmark{2}, Rahul Krishna\IEEEauthorrefmark{3},
Saurabh Jha\IEEEauthorrefmark{3}, Ravishankar K. Iyer\IEEEauthorrefmark{2}}
\IEEEauthorblockA{\IEEEauthorrefmark{2}University of Illinois Urbana-Champaign, Urbana, IL 61801, USA\;
\IEEEauthorrefmark{3}IBM Research, Yorktown Heights, NY 10598, USA}
\IEEEauthorblockA{\IEEEauthorrefmark{2}\{scui8, rkiyer\}@illinois.edu\;
\IEEEauthorrefmark{3}\{rkrsn, Saurabh.Jha\}@ibm.com}
}

\maketitle
\definecolor{bluekeywords}{rgb}{0.13, 0.13, 1}
\definecolor{greencomments}{rgb}{0, 0.5, 0}
\definecolor{redstrings}{rgb}{0.9, 0, 0}
\definecolor{graynumbers}{rgb}{0.5, 0.5, 0.5}
\definecolor{mygreen}{rgb}{0,0.6,0}
\definecolor{mygray}{rgb}{0.5,0.5,0.5}
\definecolor{mymauve}{rgb}{0.58,0,0.82}

\definecolor{peach}{HTML}{ffdda6}
\definecolor{amber}{HTML}{cc4e00}

\lstset{ %
  backgroundcolor=\color{white},   
  basicstyle=\footnotesize\ttfamily,        
  breakatwhitespace=false,         
  breaklines=true,                 
  captionpos=b,                    
  commentstyle=\color{mygreen},    
  deletekeywords={...},            
  escapeinside={\%*}{*)},          
  extendedchars=true,              
  frame=single,                    
  keepspaces=true,                 
  keywordstyle=\color{blue},       
  language=Python,                 
  otherkeywords={*,...},            
  numbers=left,                    
  numbersep=5pt,                   
  numberstyle=\tiny\color{mygray}, 
  rulecolor=\color{black},         
  showspaces=false,                
  showstringspaces=false,          
  showtabs=false,                  
  stepnumber=2,                    
  stringstyle=\color{mymauve},     
  tabsize=2,                       
  title=\lstname                   
}

\newcommand*\circled[1]{%
  \tikz[baseline=(char.base)]{
    \node[
      shape=circle,
      draw=amber,
      line width=0.9pt, 
      inner sep=1.5pt,
      fill=peach
    ] (char) {\color{amber}\scriptsize\textbf{#1}};
  }%
}
\newcommand*\circledblack[1]{\tikz[baseline=(char.base)]{
            \node[shape=circle,fill,inner sep=1.0pt] (char) {\textcolor{white}{#1}};}}

\newcommand*\circledgreen[1]{\tikz[baseline=(char.base)]{
        \node[shape=circle,draw,inner sep=1.5pt, ForestGreen, fill=ForestGreen] (char)
               {\color{white}\scriptsize\textbf{#1}};}%
        }

\newcommand*\circledred[1]{\tikz[baseline=(char.base)]{
        \node[shape=circle,draw,inner sep=1.5pt, Red, fill=Red] (char)
               {\color{white}\scriptsize\textbf{#1}};}%
        }
\definecolor{Maroon}{RGB}{192,0,0}
\definecolor{forestgreen(web)}{rgb}{0.13, 0.55, 0.13}

\newcommand\blfootnote[1]{%
  \begingroup
  \renewcommand\thefootnote{}\footnote{#1}%
  \addtocounter{footnote}{-1}%
  \endgroup
}

\begin{abstract}
Unresolved production cloud incidents cost an average of over \$2M per hour. 
This paper introduces \textbf{\agent}, an orchestrator that manages and deploys an agentic workflow for diagnosing code- and configuration-caused cloud incidents. 
\agent employs an LLM-driven structured traversal
over two types of graph: (1) a service dependency graph (\sgraph) that
captures microservice-level dependencies; and (2) a hammock-block program dependence graph (\graph) that captures code-level dependencies for each
microservice. 
Compared to state-of-the-art ReAct baselines, \agent improves RCA accuracy by up to $6.3\times$ while reducing token consumption by $5.3\times$. \agent is demonstrated on a set of 30 comprehensive real-world incidents that is being compiled into an RCA benchmark.

\end{abstract}

\vspace{-1em}

\section{Introduction}
\label{sec:intro}
Production cloud incidents cost an average of over \$2M per hour~\cite{newrelic_2025_report,newrelic_2025}. While operational mitigation actions such as reboot, restart, scale-out, and rollback can resolve many such incidents, approximately 24\% are not resolved by operational mitigation~\cite{ghosh_2022_socc}. 
In those cases, engineers supported by tools must perform \textit{root-cause analysis} (RCA), a diagnostic process that identifies the origin (i.e., root causes) of a failure.
When RCA is ineffective, outages can be prolonged and highly disruptive~\cite{monzo_2017_outage,reddit_pi_day_outage,aws_message_101925,prince_cloudflare_outage_2025}.

This paper introduces \textbf{\agent}, an orchestrator that manages and deploys an agentic workflow for diagnosing code- and configuration-caused cloud incidents. \agent employs an LLM-driven structured traversal
over two types of graph: (1) a service dependency graph (\sgraph) that
captures microservice-level dependencies; and (2) a hammock-
block~\cite{johnson_pst_1994,fubo_hammock_2004} program dependence graph (\graph) that captures code-level dependencies for each microservice. 
Together, these graphs efficiently encode microservice- and code-level dependencies.
By explicitly and jointly imposing both microservices and code dependency structures in the agentic RCA workflow, \agent extends the diagnosis beyond observability symptoms, making it capable of identifying and explaining root causes that lie in code or configurations. 
We demonstrate the breadth of \agent using 30 real-world examples, integrated into a comprehensive benchmark that is open-sourced. 

\textbf{Contribution.} Our main contributions are as follows:
\be
\item \textbf{\agent}, an agentic approach for cloud incident RCA with structured, LLM-driven graph reasoning and traversal that utilizes microservices and program dependency graphs.
\item An application of the hammock block program dependence graph for agentic RCA, leveraging the hammock block's hierarchical nesting structure to analyze microservice code at multi-granular levels seamlessly. 
\item Code-Cloud-RCA Benchmark, integrated into ITBench~\cite{itbench_k8s_topology_monitor_2025}, consisting of 30 different scenarios, each incorporating a unique software, configuration, deployment, or resource-related fault observed in the real world~\cite{cbs_database,k8saf,monzo_2017_outage,reddit_pi_day_outage,ghosh_2022_socc,gunawi2014cloudbug,hotos_19} and injected into a live Kubernetes cloud environment.

\item A demonstration of \agent's agentic capabilities to perform cross \sgraph-\graph traversal to diagnose a challenging incident.
\ee

\textbf{Results.} 
We evaluated \agent's RCA effectiveness in terms of root cause accuracy and token consumptions 
across those 30 scenarios. 
\agent achieved a root-cause reasoning accuracy of 61.5\% and a root-cause identification accuracy of 73.9\%, an increase of 6.3$\times$ and 3.4$\times$, respectively, over state-of-the-art ReAct baselines~\cite{jha2025itbench,wang2024rcagent}.
In addition, \agent reduced token consumption by $5.3\times$, from 884.9k tokens to 166.5k tokens per successful diagnosis, compared to the same baselines.

An important reason for \agent's effectiveness gain is its ability to reduce the context space and focus microservice- and code-level analysis on the regions most implicated by an incident, guided by the underlying dependency structure. 
Prior work~\cite{roy2024exploring_rca_agents,xu2025openrca,jha2025itbench,chen2025stratus,wang2024rcagent,li2025coca_icse} does not impose such a structure and relies entirely on the LLM's built-in autoregressive analysis over unstructured, plain-text prompts, which is shown to be less effective.
\agent instead explicitly constrains the LLM's reasoning along graph-defined dependencies via LLM-driven graph traversal. 

\blfootnote{\agent implementation and benchmark scenarios used in this paper are publicly available at \href{https://doi.org/10.5281/zenodo.19163486}{{{https://doi.org/10.5281/zenodo.19163486}}}.}

\vspace{-1em}

\section{Terminology}
\textbf{Root cause analysis (RCA).} 
Following the understanding espoused in~\cite{google_sre_2016}, 
RCA is the diagnostic process that
(1) identifies the microservice responsible for the incident and (2) the statement(s),
function(s), or configuration(s) identifying the origin of the incident, and (3) provides concise reasoning about its propagation and manifestation to facilitate remediation.

\textbf{Service dependency graph (\sgraph).} The service dependency graph of a cloud application is a directed graph with nodes representing individual microservices and directed edges showing the dependency between them. 
In any given interval, the graph represents the current dynamics among the microservices. 

\textbf{Program dependence graph (\graph).} A directed graph that captures microservice program structure~\cite{ferrante1987pdg}; the nodes are hammock blocks~\cite{fubo_hammock_2004,johnson_pst_1994} and the edges are data, control, and call dependencies.

\textbf{Hammock block.} 
A hammock block restructures unstructured code into a structured block corresponding to a single-entry, single-exit (SESE) region in code~\cite{fubo_hammock_2004,johnson_pst_1994}.

\textbf{Observability.} In microservices, observability is the ability to infer the distributed system’s internal state from its generated data (e.g., logs, metrics, events, and distributed traces), enabling real-time insight through analysis. 

\textbf{Code context.} 
Code context refers to code-related information.

\section{Approach Overview}
\label{sec:mot}

This section introduces \agent, an orchestrator that manages and deploys an agentic workflow for cloud incident RCA.
At the core of this approach is an LLM-driven structure traversal over two types of graph: (1) an \sgraph that captures microservice-level dependencies (providing a high-level localization of faulty microservices); and 
(2) a hammock-block \graph that captures code-level dependencies for each microservice (allowing for fine-grained RCA decisions that identify offending code paths and configurations responsible for the incident). 
\agent's hammock-block \graph provides an efficient structural representation of a microservice program, enabling seamless analysis of microservice code at different granularities.
By integrating tools, observability data, and dependency graphs, \agent delivers demonstratively accurate RCA for live cloud incidents, demonstrated on microservice-based Kubernetes applications. 
In summary, the core  functionalities of \agent's workflow include the following.

\textbf{Data gathering.} 
Like prior agentic approaches~\cite{roy2024exploring_rca_agents,jha2025itbench,xu2025openrca,chen2025aiopslab,chen2025stratus,wang2024rcagent,li2025coca_icse}, \agent uses cloud monitoring tools~\cite{prometheus_2025,jaeger_2025,clickhouse_2025} to actively collect a set of automated and user-specified alerts referred to as \textit{golden-signal alerts} (e.g., error-rates or latencies exceeding user-defined thresholds) and observability data consisting of cloud application and microservice logs, distributed traces of microservice calls, Kubernetes events, and application metrics.
\agent also actively retrieves microservice topology, as in~\cite{wu_microrca_2020,2021_sage}, using a cloud topology monitoring platform~\cite{datadog_2025}, which is used to build the \sgraph.
In addition, \agent uses microservice codebases for constructing \graph for microservices.

\textbf{\sgraph and \graph construction.} As the key differentiator from prior work~\cite{wang2024rcagent,li2025coca_icse}, \agent's agentic workflow utilizes two modalities of graphs. 
The first is a dynamically evolving service dependency  graph~\cite{wu_microrca_2020,2021_sage} for the cloud application, with currently deployed microservices as nodes and inter-microservices dependencies as edges, generated via a real-time topology monitor~\cite{datadog_2025}. The second is   
a hammock-granularity \graph for each microservice node in the \sgraph, with hammock blocks as nodes and control, data, and function-call dependencies as edges. This \graph is generated by static program analysis tools~\cite{tree_sitter_zenodo_2025,rahul2024cldk} and represents the software dependency of a microservice application relevant to root-cause analysis.

Together, the \sgraph and \graph capture relevant microservice-level and code-level dependencies for RCA.
These graphs serve as bases for RCA via LLM-driven graph traversal, first at the microservice level and subsequently at the code level, to identify responsible microservices, faulty code sites, and/or configurations that might explain the observed erroneous signals (such as active alerts, error logs, traces, metrics, and events), providing in-depth RCA down to the code level.

\textbf{Structure-aware agentic RCA.}
After being triggered by active golden-signal alert(s), \agent presents the relevant data (active alert(s) and error traces) to 
an LLM to identify and select  microservices correlated with error signals in those data. 
The selected microservices become initial candidates for agentic focus.
Subsequently, the LLM investigates each microservice by performing structure-aware reasoning through traversal of the corresponding hammock blocks. 
\agent uniquely constructs the \graph using hammock blocks at different granularities spanning module, class, function, and statement levels, allowing the LLM to seamlessly move between levels as required to optimize analysis efficiency and accuracy.
Leveraging this structure, the LLM iteratively traverses first the \sgraph and then the hammock-block \graph at the required level to narrow down the pool of potentially responsible microservices and the corresponding fault sites in code or configuration, until it reaches a conclusive decision on the root cause. 

Unlike prior agentic approaches that process code as monolithic text~\cite{wang2024rcagent,li2025coca_icse} for RCA, \agent provides an efficient approach for strategically traversing the \sgraph and the corresponding hammock blocks in \graph, analyzing local code context alongside observability data. 
The graph traversal constrains the LLM's diagnosis to incident-relevant dependency paths and accumulates code insights that could explain the observed error signals (e.g., alerts, error traces, error logs), while filtering out irrelevant paths. 
Thus, instead of relying solely on the
LLM's attention mechanism to implicitly infer code structure and focus on context relevant to the incident,
\agent's workflow explicitly imposes structure-aware, concise, and accurate diagnosis of the underlying incident.
Based on the observability data and the accumulated code insights,
the LLM classifies the microservice under investigation as \textsc{Primary Failure}, \textsc{Symptom Only}, or \textsc{Unrelated}. 
Upon finishing its investigation of the current microservice, \agent moves on to investigate the other selected microservices and the dependees according to the \sgraph.
Once all initially selected microservices and their dependees have been investigated, \agent calls the LLM to consolidate the per-microservice investigation decision into a final RCA report detailing the fault's origin, propagation, and impact.

\begin{figure}[t!]
    \centering
    \includegraphics[width=1.0\linewidth]{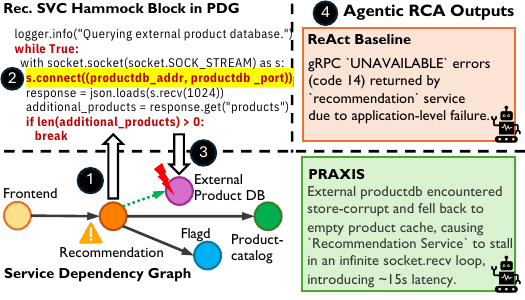}
    \caption{\label{fig:praxis_motivation} \textbf{Incident:} Degraded external database
returned empty responses, triggering a silent retry loop in the Recommendation service that manifested solely as a high-latency alert associated with the Recommendation service, without explicit error logs or error traces.
\textbf{Cross-\sgraph-\graph traversal:} (1) LLM selects the Recommendation service for investigation based on the observed alert. (2) Investigation of the Recommendation service reveals a silent retry loop with missing error logs. Code traversal shows that an unresponsive External Product Database triggers this loop. (3) Subsequent investigation shows a storage failure in the External Product DB as the root cause. (4) The ReAct agent baseline fails to pinpoint the precise root cause, whereas \agent successfully isolates the root cause. 
}
\end{figure}

\textbf{Cross-\sgraph-\graph traversal: a challenging scenario.} A key innovation of our approach is the ability to handle incidents for which multiple microservices are identified as potential root causes. 
In such cases, the LLM must traverse from the first microservice into its specific hammock block within the \graph, and then transition to a second microservice, according to the \sgraph and its respective hammock blocks, to form a coherent and holistic RCA decision. 
\agent facilitates this by attaching the \graph of each microservice to the global \sgraph, ensuring that the LLM can perform these cross-layer traversals when needed. 
We demonstrate the necessity of this unique capability in \Cref{fig:praxis_motivation}, using a challenging incident that the baseline ReAct agent failed to solve.

\section{\agent: Methodology}
\label{sec:met}
\begin{figure}[t!]
    \centering
    \includegraphics[width=0.99\linewidth]{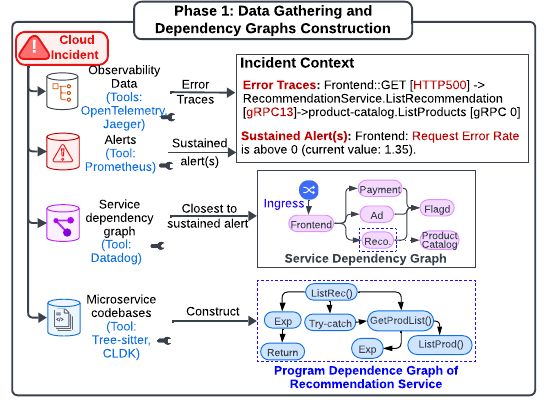}
    \caption{\agent Phase 1: Data gathering and dependency graph construction. 
    }
    \label{fig:praxis-phase-1}
\end{figure}

This section details the methodology of \agent, which employs an iterative reasoning process to perform RCA on cloud incidents. \agent operates in the following phases:

\noindent\textbf{Phase 1 (\tion{datagathering},~\Cref{fig:praxis-phase-1}): Data gathering and dependence graph construction.}~\agent collects the \textit{incident context}: error traces from Jaeger~\cite{jaeger_2025}, sustained alerts from Prometheus~\cite{prometheus_2025}, the \sgraph snapshot from the cloud topology monitoring platform~\cite{datadog_2025}, and the \graph, constructed using Tree-sitter~\cite{tree_sitter_zenodo_2025} and CLDK~\cite{rahul2024cldk}, from the microservice source code.

\noindent\textbf{Phase 2 (\tion{bootstrap},~\Cref{fig:praxis-phase-2}): Microservice candidate(s) selection.}~ 
Using the incident context (\ie, error traces and sustained alerts), \agent employs an LLM to identify initial \textit{root-cause candidate} microservices.

\noindent\textbf{Phase 3 (\tion{rootcause},~\Cref{fig:praxis-phase-3,fig:praxis-pdg-trav}): RCA decision-making.} 
\agent directs an LLM to iteratively traverse the microservice’s \graph to construct \textit{program context} and diagnose code regions relevant to the incident, then assign an RCA judgment (\textsc{Primary Failure}, \textsc{Symptom Only}, or \textsc{Unrelated}) before advancing to the next microservice as suggested by the \sgraph.

\noindent\textbf{Phase 4 (\tion{summary},~\Cref{fig:praxis-phase-4}): Final RCA summary.}~Upon completing all investigations, \agent consolidates the LLM’s judgments and reasoning to generate a comprehensive RCA report (\Cref{fig:praxis-phase-4}) that provides (1) root-cause identification and (2) root-cause reasoning down to the responsible statements, functions, or configurations.

\subsection{Phase 1: Data Gathering\label{sec:datagathering}}
In Phase 1 (\Cref{fig:praxis-phase-1}) of the agentic RCA workflow, \agent leverages cloud monitoring tools~\cite{prometheus_2025,jaeger_2025,clickhouse_2025} to actively monitor alerts (e.g., error rates, latency). 
Upon detecting a sustained alert indicating a sustained SLO violation, it collects and aggregates the relevant observability data to construct the \textit{incident context}, as detailed below.

\be
\item \textit{Alerts.} Alerts 
signal service-level objective (SLO) violations, including high latencies and elevated error rates.  
\agent continuously polls the Prometheus API for any active sustained alerts; these alerts serve as the initial triggers for \agent's RCA workflow.

\item \textit{Distributed error traces.}
Distributed error traces, \REV{obtained via cloud-native observability tools (e.g., OpenTelemetry and Jaeger)},  record runtime request latencies and propagations of inter-service requests, explicitly highlighting error dependency paths across microservice boundaries \REV{for initial fault localization.}

\item \textit{Logs.}~Cloud application and microservices log entries and their associated error/warning patterns\footnote{To avoid context overflow, only the $K$ most recent entries are collected. We set $K=40$ for our evaluation, as empirical testing showed that increasing the value further does not significantly alter RCA accuracy.} 
provide rich runtime context (e.g., information, warnings, errors, stack traces, exceptions), offering direct insight into the timing and nature of runtime executions, errors, and failures. 
\item \textit{Events.}~Kubernetes events regarding deployment updates and lifecycle changes that affect the microservice or its resources (\eg, ReplicaSet scaling, container restarts) provide diagnostic context on deployment/resource configuration changes that could reveal the root cause. 
\item \textit{Metrics}~include aggregated error and warning counts derived from logs and events, alongside span and trace latency measurements to quantify the severity of observed anomalies.

\ee

\agent aggregates sustained alerts and error traces into the incident context, which serves as the initial description of the cloud incident. Moreover, logs, metrics, and events will be used in Phase 3 for per-microservice investigation. 

\Cref{fig:praxis-phase-1} illustrates an example incident context for a cloud incident in which the Frontend service is unable to recommend products: (1) a sustained \texttt{Request Error Rate} alert is observed in the Frontend service, and (2) the corresponding error traces reveal \texttt{gRPC13} errors during calls to the \texttt{Recommendation.ListRecommendation} endpoint.

\noindent\textbf{\sgraph \& \graph}
\agent's agentic workflow utilizes two graphs: 
(1) a dynamically evolving \sgraph for the cloud application, representing the current microservices configuration that is being deployed and executed; and  
(2) a hammock-block \graph for each microservice node in the \sgraph, representing the software dependency of a microservice program relevant to root-cause analysis. 
Together, these graphs capture relevant microservice-level and code-level dependencies and serve as bases for agentic RCA via LLM-driven graph traversal.
\agent constructs these graphs in Phase 1 after gathering data, as shown in~\Cref{fig:praxis-phase-1}.

\paragraph{Service dependency graph\label{sec:sdg} (\sgraph)} 
\agent employs a \textit{service dependency graph} (\sgraph) to capture the currently deployed microservices and their dependencies at the time of the incident. A \emph{\sgraph} models the dependency structure of a cloud application
\(C = \{e_1, \ldots, e_n\}\), where each \(e_i\) is a microservice or one of its
associated resources (pod or ConfigMap)~\cite{wu_microrca_2020,2021_sage}. It is
defined as a directed graph \(G_C = (V, E_{\text{depend}}, \lambda_V)\) with
\(V = C\), where an edge \((u, v)\) indicates that microservice \(u\) depends on
\(v\) (another microservice or one of its resources), and non-microservice nodes
serve as leaves with no outgoing edges. Each node is annotated by \(\lambda_V\)
to record its kind and name, completing the definition of the \sgraph.
The \sgraph $G_C$ is constructed and maintained by a cloud topology monitor~\cite{datadog_2025}. 
To capture the dynamic nature of the cloud environment, the graph is updated at predefined intervals.
Consequently, \agent retrieves the graph snapshot at the time 
of the alert to bootstrap the RCA process.
Since the investigation can extend to non-microservice cloud components such as a pod or a configMap associated with a microservice, we hereinafter refer to a node in the \sgraph as an \textit{entity}.
\Cref{fig:praxis-phase-1} presents an example \sgraph\footnote{For brevity, we do not show all nodes/edges in the figure.} of the cloud application~\cite{opentelemetry_demo} employed by our benchmark. 

\paragraph{\label{sec:pdg}Program dependence graph (\graph)}
To perform fine-grained diagnosis, \agent augments each microservice node in the \sgraph with a \textit{program dependence graph} (\graph)~\cite{ferrante1987pdg}, enabling reasoning over code paths and potential fault sites. For a microservice program \(P\) comprising a finite set of single-entry–single-exit (SESE) hammock blocks \(B = \{b_1, \ldots, b_n\}\)~\cite{johnson_pst_1994,fubo_hammock_2004}, the \graph is defined as a labeled directed graph \(G_P = (V, E, \lambda_V, \lambda_E, \prec)\). Here, \(V = B\), and edges \(E = E_{\text{ctl}} \cup E_{\text{data}} \cup E_{\text{call}}\) capture control-, data-, and call-dependence between blocks. \(\lambda_V\) and \(\lambda_E\) annotate nodes and edges with their corresponding semantic attributes: 
\REV{\(\lambda_V\) includes node type, defined and used variables, string literals, and code snippets, and \(\lambda_E\) includes dependency type (control-flow,
data-flow, or caller-callee) and associated variable or function names.}
A containment relation \(\prec\) specifies when one hammock block syntactically contains another, inducing a hierarchical structure over module-, class-, function-, and statement-level regions. This hierarchy allows \agent to traverse and reason about the program at the most informative granularity for RCA.

Provided with the microservice codebases, \agent automatically generates the \graph for microservices as part of the initialization process, using static program analysis tools: Tree-sitter~\cite{tree_sitter_zenodo_2025} for parsing hammock blocks and CLDK~\cite{rahul2024cldk} for data, call, and control relationship analysis.
The PDGs generated are stored as adjacency lists in JSON files. This ensures a lightweight, language-agnostic universal format that facilitates efficient storage, management, update, and query.
During agentic RCA, \agent loads the \graph for each microservice under investigation at runtime.

\Cref{fig:praxis-phase-1} shows a representative \graph of the Recommendation service from~\cite{opentelemetry_demo}, comprising function- and statement-level hammock blocks.\footnote{We omit the full \graph for brevity, as the actual structure contains 44 hammock blocks.} 
\texttt{Exp} denotes expression statements (blocks of contiguous, straight-line statements without branching), whereas other nodes represent branching logic (e.g., the \texttt{try-catch} node). 
Function-level blocks are explicitly labeled with function names (e.g., the \texttt{GetProdList()} node).

\subsection{Phase 2: Microservice Candidate(s) Selection\label{sec:bootstrap}}

\begin{figure}[t!]
    \centering
    \includegraphics[width=1.0\linewidth]{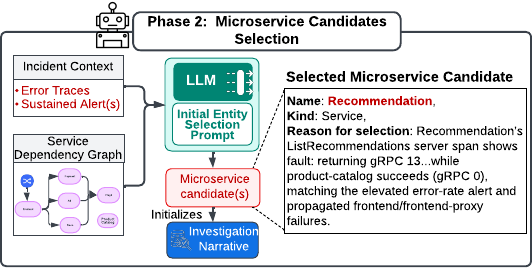}
    \caption{\agent Phase 2: Initial microservice candidate(s) selection.}
    \label{fig:praxis-phase-2}
\end{figure}

In Phase 2 (\Cref{fig:praxis-phase-2}), \agent invokes an LLM to select an initial set of cloud \textit{microservice candidates} potentially responsible for the incident to start the investigation process using the \sgraph and the incident context from Phase 1.
\agent prompts the LLM (denoted by \mathllm{}) with $\psi_{{select}}$, conditioning the selection on the \sgraph $G_{C}=\left(V, E\right)$ and the incident context, which comprises active sustained alert(s) ($A$) and collected error traces ($T$).
The LLM suggests a set of initial microservice candidates (up to $N$\footnote{$N$ is a user-defined variable that we set to $N=5$ for our experiments. Further increases of this value yielded no improvements to our results.}) from nodes in $G_C$, and outputs an initial \textit{investigation narrative} $I_0$ describing the observed alerts, traces, and the reasons why they were selected.
In addition, \agent initializes an investigation queue $Q$ to actively keep track of microservices that need to be investigated and initializes $Q$ with those selected candidates: $Q \leftarrow \mathcal{M}(\psi_{{select}}(A, T, \lambda_{V}(G_C)))$.

\Cref{fig:praxis-phase-2} shows an example microservice candidate selection, based on the sustained alert and error traces from Phase 1. LLM selects the Recommendation service as an initial candidate, noting that \texttt{gRPC} calls from the Frontend service exhibit \texttt{gRPC 13} errors. These errors propagate upstream, manifesting as \texttt{HTTP500} errors observed in the Frontend service.

The initial investigation narrative $I_0$ and the investigation queue $Q$ serve as the starting point for \graph traversal and RCA decision-making in Phase 3.

\begin{figure*}[t!]
    \centering
    \includegraphics[width=1.0\linewidth]{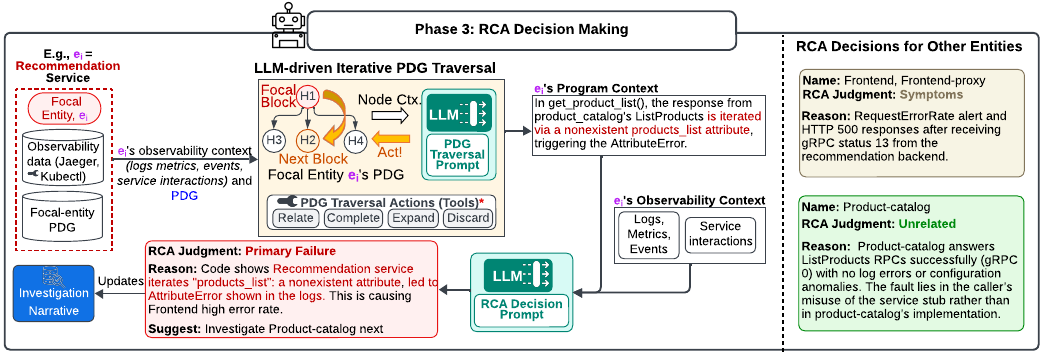}
    \caption{\agent Phase 3: RCA decision-making. 
    This process is repeated for the next focal entity that is (1) a dependee of the current focal entity and/or (2) suggested by the LLM based on the focal entity's RCA decision. 
    }
    \label{fig:praxis-phase-3}
\end{figure*}
\subsection{Phase 3: RCA Decision-making \label{sec:rootcause}}
In Phase 3 (\Cref{fig:praxis-phase-3}), given the investigation queue $Q$ and the initial investigation narrative $I_0$ from Phase 2, \agent iteratively dequeues and investigates an entity $e$ from the head of $Q$ and determines its role: a root cause, a symptom, \REV{or unrelated to} the cloud incident.

For each dequeued entity $e_i \in Q$ (designated as the \textit{focal entity}) at iteration $i$, \agent first constructs the \textit{observability context} $c_i$ and retrieves the corresponding \graph $G_P$.
Crucially, \agent then employs an \emph{LLM-driven \graph traversal} that traverses the entity's ($e_i$'s)~\graph ($G_P$) to construct an incident-centric program context $C_P$ that explains the observed error symptoms from the perspective of entity code context.
Given the observability context $c_i$ and program context $C_P$ of the focal entity, \agent uses the LLM to determine its role in the cloud incident as one of  \textsc{Primary Failure} (the root cause), \textsc{Symptom Only} (a symptom), or \textsc{Unrelated}.
After finishing the investigation of the current entity, \agent moves on to investigate the other LLM-selected
microservices and the dependees according to the \sgraph $G_C$ by adding them to the investigation queue $Q$. 
\agent repeats this analysis on all entities in $Q$ until $Q$ is empty.

Specifically, this phase consists of the following steps:

\noindent\textbf{(a)~Observability context construction.}~\agent first retrieves the focal entity $e_i$'s observability data, comprising logs, metrics, and events, the \sgraph ($G_C$), and the focal entity's \graph ($G_P$), built in Phase 1.  
\agent additionally collects inter-service interactions (e.g., inbound/outbound calls) as indicated by the \sgraph and physical infrastructure attributes (e.g., cluster, node IDs) for each microservice-type entities.
It also collects the current configuration status using \texttt{Kubectl} for ConfigMaps-type entities.
\agent structuralizes these collected data into a JSON representation
at the time of incident investigation for the LLM to consume as \textit{observability context} $c_i$, providing the most up-to-date runtime evidence to anchor the agentic program analysis and RCA reasoning.

\noindent\textbf{(b)~LLM-driven \graph traversal.\label{sec:traver}}~Leveraging the observability context $c_i$ and the graph $G_P$, the LLM agent operates in an iterative loop to traverse code paths defined by hammock blocks and code dependencies in the \graph. It analyzes how the underlying code logic aligns with the observed signals to pinpoint the root cause and explain the incident.

\noindent\textbf{Initial hammock block (\graph node) selection.}~
The first step to ensure that the \graph traversal begins from a semantically relevant point tied to the observed cloud incident is to identify the initial node $b_{0}$ in the \graph $G_P$ as the starting point of the traversal.
This node represents the code region most relevant to the observed symptom described in the entity's observability context $c_i$ (e.g., error descriptions, exception stack-trace, file name, and line numbers).

Given $c_i$ and the \graph, 
we employ a language model \mathllm{} conditioned on a prompt composition function $\psi_{match}$ that encodes the runtime observability context $c_i$ and the node attributes of the \graph $\lambda_{V}$ (see~\tion{pdg}). Specifically, $\psi_{match}\left(c_i, \lambda_V\right)$ constructs a joint representation with:
\bi
\item[(1)] observability data such as error logs, exception stack frames, error events, and metric anomalies;
\item[(2)] \graph node metadata including the method names, corresponding code fragments, and file/line identifiers and string literals.
\ei

The language model \mathllm{} then evaluates the correspondence between those two modalities and selects an initial hammock block node from \mathpdg:
\begin{equation*}
    b_{0} = \mathcal{M}\left(\psi_{match}\left(c_i, \lambda_V \in V(G_P)\right)\right) 
\end{equation*}
where $b_{0}\in V(G_P)$ is a hammock block whose attributes (e.g., a string literal) correlate to the observed signals (e.g., a log entry).  
\agent defaults to the hammock block corresponding to the entry-point (e.g., the \texttt{main} function) or request-handling methods (e.g., an API-handler function) of the microservice program when observability data are absent or insufficient due to the incident or when the matching is ambiguous (e.g., multiple matches).

\noindent\textbf{\graph traversal.}~%
Once the starting hammock block $b_0$ (i.e., a node in \graph) has been identified, \agent orchestrates the LLM's iterative agentic traversal of the \graph (\mathpdg) to construct program context $C_P$ for in-depth RCA. 
Here, the program context $C_P$ summarizes the microservice code paths that are likely related to the incident and explains the observed error symptoms from the perspective of this code path.
As such, $C_P$ provides rich, in-depth diagnostic context beyond symptom-level observability context for determining the microservice's role in the incident. 

We define this agentic process as a 4-tuple:

$\qquad\qquad\qquad\qquad\mathcal{A} = \left<\mathcal{M}, \mathcal{T}, \mathcal{S}, G_P\right>$
 
\bi
\item $\mathcal{M}$ is the LLM used for reasoning and decision-making;
\item $\mathcal{T}$ is the tool-set used by $\mathcal{M}$ to traverse over the \graph; 
\item $\mathcal{S}$ is the agent's state space;
\item $G_P$ is the \graph for microservice $e_i$ that acts as a transition function according to the tool action. 
\ei

The PDG traversal starts from the initial hammock-block node, $b_{j}=b_0$, which serves as the initial anchor in the PDG. At each \graph-traversal iteration $j$, we define the state as: 

$\qquad \qquad s_j= \left<b_j, B_{related}, c_i, H_j\right> \in \mathcal{S}$ 

\noindent where $b_j\in V(G_P)$ is the current hammock-block node under analysis; $B_{related}$ is the set of blocks in $G_P$ related to $b_j$ by control, data, or call dependencies ($B_{related} = \{b | (b_j, b_k) \in E(G_P)\}$); $c_i$ is the current microservice's observability context; and $H_j$ is the past traversal history up to step $j$ as \graph-traversal memory.

Conditioned on the current state $s_j$, the LLM $\mathcal{M}$ produces a code-insight ($\iota_j$) and the graph-traversal action ($a_j$) pair, \ie, $\left(a_j, \iota_j\right) = \mathcal{M}\left(\psi_\tau(s_j)\right)$, where $\mathcal{M}$ is the LLM executing the reasoning and policy selection. 
$\psi_\tau$ is the prompt composer that encodes the current state $s_j$ into a structured form parametrized by a template $\tau$. In particular, $b_j$ and $B_{related}$'s corresponding code snippets and relations are fetched by $\psi_\tau$.
$a_j\in \mathcal{T}$ is the next chosen action (mapped to our tools),
and $\iota_j$ is the natural-language insight derived as the outcome of the current state/action. 
Notice that the LLM only consumes code context local to hammock blocks ($b_j$, $B_{related}$) for each traversal step, instead of large code files or codebases; this reduces the reasoning context space and allows the LLM to focus on concise, accurate context relevant to the incident, as indicated by $c_i$.

The agent's traversal action is implemented with a set of PDG traversal tools
{\small$\mathcal{T}$}. 
Each tool in $\mathcal{T}$ corresponds to a concrete, admissible traversal action on the PDG \mathpdg.~\Cref{tab:pdgtools} provides a detailed summary of the above tools. 
When an inadmissible action (e.g., a nonexistent block) is rejected by $\mathcal{T}$, \agent prompts the LLM to regenerate a traversal action, subject to a predefined retry limit.
Each action invocation updates the traversal history ($H_{i+1}$) and fetches the next node ($b_{j+1} \in G_P$) deterministically based on $a_j$ via 
$\mathcal{T}$ on $G_P$:
$b_{j+1} = \mathcal{T}(G_P; a_j),\; H_{j+1} = H_j \cup\{b_j, \iota_j, a_j\}$; we can then construct $s_{j+1}$ accordingly for the next iteration.

\begin{figure}[t!]
    \centering
    \includegraphics[width=1.0\linewidth]{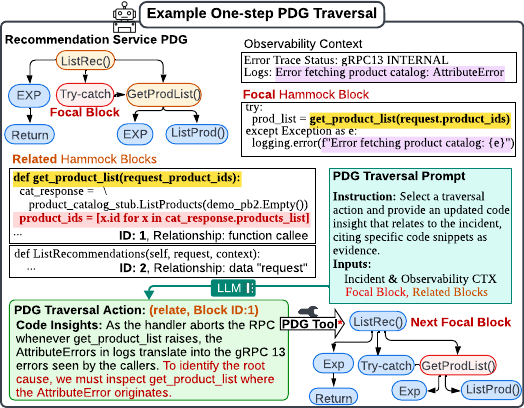}
    \caption{Example LLM-driven \graph traversal.}
    \label{fig:praxis-pdg-trav}
\end{figure}

\Cref{fig:praxis-pdg-trav} illustrates a \graph traversal step in which the LLM selects the \texttt{Relate} action to focus on Block 1 (the \texttt{get\_product\_list} function) for further investigation of the observed \texttt{AttributeError}. The LLM reasons that although the error is caught in the current focal block (the \texttt{try-catch} statement), it likely arises during execution of \texttt{get\_product\_list}.

The traversal terminates when one of the actions, \textsc{Complete} or \textsc{Discard}, is produced, when all the nodes are exhausted, or when the user-defined budget (e.g., max number of blocks) is reached. Upon completion, the agent updates its memory state $H$ by committing the final traversal record, yielding the complete traversal history $H_T$ (assuming $T$ traversal steps in total).
Notice that the visited code blocks ($\{b_0, b_1, ...,b_T\}$) encompassed in $H_T$ form a code dependency path admitted by control, data, and call dependencies.
Then, the LLM uses a prompt $\psi_{syn}$ to combine traversal history and observability context, producing the final program context $C_p(e_i)$ for microservice $e_i$:
$C_P(e_i) = \mathcal{M}(\psi_{syn}(H_T, c_i))$.
Here, $C_P$ highlights code regions, dependencies, and explanations, which link error signals in the observability data with program evidence (code snippets and dependency flows) that might explain the incident. 

\begin{table}[t]
    \centering
    \caption{\label{tab:pdgtools}PDG traversal tools available to the agent during graph-based program analysis.}
    \resizebox{\linewidth}{!}{
    \begin{threeparttable}
    \begin{tabular}{l|c|>{\raggedright\arraybackslash}p{3.0cm}|>{\raggedright\arraybackslash}p{3.0cm}}
        \hlineB{2}
        \rowcolor[HTML]{EFEFEF}
        \textbf{Tool} & \textbf{PDG Operation} & \textbf{Description} & \textbf{Effect on Reasoning} \bigstrut\\
        \hlineB{2}
        \textbf{Expand}   
            & parent($b_i$)   
            & Moves to the immediate parent or dominator node.  
            & Broaden to examine higher-level hammock blocks or callsites. \bigstrut\\\hline
        \textbf{Relate}   
            & $\Gamma(b_i)$   
            & Visits neighboring nodes linked by call, control, or data dependencies.  
            & Explores adjacent code paths to collect supporting or refuting evidence. \bigstrut\\\hline
        \textbf{Complete} 
            & ---              
            & Terminates traversal and triggers synthesis of program context $C_P$.  
            & Consolidates and finalizes the code-level hypothesis. \bigstrut\\\hline
        \textbf{Discard}  
            & ---              
            & Terminates traversal without using current findings.  
            & Prunes irrelevant or low-confidence trajectories. \bigstrut\\\hlineB{2}
    \end{tabular}
    \begin{tablenotes}[para]
        \footnotesize
        The traversal tool-set $\mathcal{T}=\{\textsf{Expand}, \textsf{Relate}, \textsf{Complete}, \textsf{Discard}\}$ defines the discrete action space of the agent.  
        Each action is implemented as a callable operation on the versioned PDG stored in the Neo4j-like graph database.
    \end{tablenotes}
    \end{threeparttable}
    }
\end{table}

\noindent\textbf{(c)~Entity role judgment and queue update.\label{sec:entity_role}}~The final stage in this phase is entity role judgment and an investigation queue update, in which \agent provides a diagnostic judgment on the entity $e_i$'s role in the incident and an admissible explanation with evidence, and updates the investigation queue accordingly. 

\noindent\textbf{Entity role judgment.}
After the program context for the entity $e_i$ has been synthesized, \agent determines the role of the current entity $e_i$ in the incident: whether $e_i$ is the root cause (primary cause of failure), a symptom, or unrelated.
The role is inferred by a language model $\mathcal{M}$ conditioned on a prompt $\psi_{j}$ that composes observability ($c_i$) and program context ($C_P$) into a reasoning prompt for the LLM $\mathcal{M}$, \ie, $J = \mathcal{M}(\psi_j(c_i, C_P))$, where $J$ is one of \{\textsc{Primary failure}, \textsc{Symptom only}, or \textsc{Unrelated}\}.
The model returns both the categorical judgment $J$ and its natural language reasoning ($I_{e_i}$) with evidence grounded in both observability and program context ($c_i$ and $C_P$). $I_{e_i}$ is then appended to the global investigation narrative $I = I \cup I_{e_i}$ used in subsequent investigation.

\noindent\textbf{Investigation queue update.}~The entity investigation queue $Q$ is then expanded to include new entities that are (1) dependees of the current focal entity, and/or (2) suggested by the LLM based on the focal entity's RCA decision.
For (1), \agent enqueues all unvisited dependees of $e_i$ according to the \sgraph $G_C$, denoted by $Q_{G_C} = G_C.\texttt{getDependees}(e_i)$, where each dependee is a cloud entity on which $e_i$ depends.
For (2), \agent prompts the LLM $\mathcal{M}$ using the entity-queuing prompt $\psi_Q$ to identify up to $k$ additional entities\footnote{We set \(k=3\) in our implementation; however, across all evaluation runs, the LLM selected at most two entities.} from $G_C$. This selection is based on the judgment $J$, the observability context $c_i$, the program context $C_P$, and the nodes in the \sgraph $G_C$, such that $Q_{\mathcal{M}_Q} = \mathcal{M}(\psi_Q(J, c_i, C_P, G_C))$.
Finally, the queue is updated: $Q = Q \cup Q_{\mathcal{M}_Q} \cup Q_{G_C}$.
Mechanism (2) is crucial as it allows the LLM to investigate entities in $G_C$ that surface from the program context $C_P$ despite having partial or missing observability data (due to insufficient coverage or the incident itself). Our evaluation confirms that this capability helps diagnose entities that do not emit observability data (e.g., a ConfigMap) or have failed silently.

\Cref{fig:praxis-phase-3} depicts the investigation of the Recommendation service (selected in Phase 2,~\Cref{fig:praxis-phase-2}), which was classified as a \textsc{Primary Failure} based on its program ($C_P$) and observability ($c_i$) contexts as well as subsequent RCA judgments for the Frontend, Frontend-proxy, and Product-catalog services.

\agent \textbf{repeats} this phase (Phase 3) for the next entity $e_{i+1}$ at the head of the updated queue $Q$. By cross-traversing between the \sgraph and \graph, \agent can resolve cloud incidents that require multiple hops to root-cause the incident while maintaining a coherent RCA summary via a consistent stream of investigation narrative $I$, a feature that is unobtainable by simply combining independent agents~\cite{llm_rca_2023,chen2025stratus,xu2025openrca,sweagent_2024}.
Critically, leveraging code-level insights to bypass gaps in the \sgraph, \agent can traverse across missing links (e.g, error traces, logs) in observability data, as shown by the Cross-SDG-PDG traversal example in~\tion{mot}.

Moreover, by iteratively traversing the \graph node-by-node, \agent progressively refines the RCA scope, transitioning from broader service-level observability signals to a more focused analysis of specific code regions for each focal entity. 
Consequently, at each step, the LLM is strictly exposed to the local context of the current graph node rather than multiple microservices or codebases/code files. 
This isolation of unrelated information (other node contexts) improves context efficiency, reduces token consumption, and mitigates potential context overflow, allowing \agent to scale effectively to large cloud applications and codebases.

\begin{figure}[t!]
    \centering
    \includegraphics[width=1.0\linewidth]{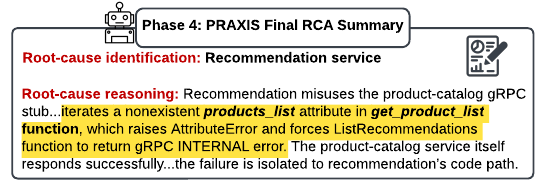}
    \caption{\agent Phase 4: Final RCA summary.}
    \label{fig:praxis-phase-4}
\end{figure}

\subsection{Phase 4: Final RCA Summary}
\label{sec:summary}
After all entities in $Q$ have been investigated, the LLM $\mathcal{M}$ synthesizes a comprehensive RCA report using its accumulated investigation history $I$ from Phase 3 with a prompt $\psi_{summary}$ to produce a diagnosis report ${D}$, \ie, $D = \mathcal{M}(\psi_{summary}(I))$. 
$D$ identifies the primary failure entities (root-cause entities), root-cause reasoning, and fault propagation chain, while providing a justification for each entity's diagnosis, citing evidence from the microservice code and observability data. This step consolidates the investigation and generates a structured, human-readable RCA report grounded in both observability and program-level reasoning.

\Cref{fig:praxis-phase-4} presents an example final RCA report. \agent correctly identifies the Recommendation service as the root cause, specifically attributing the failure to a schema mismatch (accessing nonexistent fields), and suggests remediation. It also accurately classifies Frontend and Frontend-proxy as \textsc{Symptom Only} and Product-catalog as \textsc{Unrelated}.

\section{Experiment Setup}
\label{sec:exp}
This section first clarifies the problem and evaluation scope, and then presents the cloud incident scenarios, baselines, and metrics used for evaluation.

\subsection{Problem and Evaluation Scope}
We evaluate \agent on application-level incidents, focusing on microservice codebases for which source code is available for analysis. 
Our approach is beneficial even under partial code availability, as reasoning over available code improves diagnostic accuracy (see \tion{baseline}); when source code is unavailable, \agent falls back to using only observability and incident contexts for RCA. 
Gains are naturally limited when no code is available (see \tion{ablation}). 
\agent's scope is confined to incident diagnosis within the investigation/diagnosis phase of the cloud incident lifecycle~\cite{google_sre_2016,google_cloud_incident_lifecycle}.

\subsection{Cloud Incident Evaluation Scenarios}
We developed the \textit{Code-Cloud-RCA Benchmark} comprising 30 cloud incidents using~\cite{opentelemetry_demo}, the same cloud application used  in~\cite{chen2025aiopslab}, supported by a standard monitoring stack~\cite{jaeger_2025,prometheus_2025,clickhouse_2025,opentelemetry_overview,datadog_2025}.
These scenarios are distilled from root-cause types commonly reported in production studies~\cite{cbs_database,k8saf,hotos_19,gunawi2014cloudbug,ghosh_2022_socc,yan2025empirical} and real-world incidents~\cite{k8saf,marcomutiny2024}. 
Table~\ref{tab:incident-typologies} enumerates the scenario templates, fault mechanisms, expected symptoms, and reference to real-world incidents; the rightmost column (\#) records the number of scenarios per template. 
As with any benchmark, comprehensive coverage cannot be guaranteed, and by providing this benchmark, we invite the broader community to contribute to and expand its scope.  
Scenarios driven by the same template vary in timing (startup vs. serving), observability (stack traces vs. error logs vs. error codes), failure modes (stalls vs. rapid retries), or configurations, and are annotated with ground-truth root-cause microservices and reasoning. 
\begin{table}[t]
    \centering
    \caption{\label{tab:incident-typologies}
    Incident scenario descriptions and real-world reference.
    Deployment/resource cases are included as negative examples to highlight \agent's limitations.
    }
    \begin{threeparttable}
    {\fontsize{8}{10}\selectfont
    \setlength{\tabcolsep}{3pt}
    \renewcommand{\arraystretch}{1.1}
    \begin{tabularx}{\columnwidth}{%
        >{\raggedright\arraybackslash}p{0.24\columnwidth}|%
        Y|%
        >{\raggedright\arraybackslash}p{0.32\columnwidth}|%
        >{\centering\arraybackslash}p{0.05\columnwidth}}
        \hlineB{2}
        \rowcolor[HTML]{EFEFEF}
        \textbf{Scenario description} & \textbf{Mechanism / How simulated} & \textbf{Typical symptoms (Real-world incidents)} & \textbf{\#} \bigstrut\\
        \hlineB{2}

        Data schema mismatch/incompatibility between services
            & Access a nonexistent field (\texttt{products\_list}) in a protobuf response from another microservice.
            & Unhandled exception; 5xx bursts; request failures. (Monzo 2017 outage~\cite{monzo_2017_outage}; report~\cite{hotos_19,ghosh_2022_socc,k8saf})
            & 4 \bigstrut\\\hline

        Improper handling of external dependency failures
            & Upstream DB/service hangs or returns empty results; dependent service stalls without timeout or enters rapid retries.
            & Latency spikes; retry storms; high error rate. (Discord 2023 outage~\cite{discord2023outage}; report~\cite{cbs_database,gunawi2014cloudbug,hotos_19,ghosh_2022_socc})
            & 6 \bigstrut\\\hline

        Internal logic bug
            & Invert an assertion or suboptimal implementation (e.g., an exponential-time LCS ranking algorithm).
            & Slowness/timeouts; intermittent crashes; incorrect results. (Cloudflare 2019 outage~\cite{grahamcumming2019cloudflare}; report~\cite{cbs_database,gunawi2014cloudbug,hotos_19,yan2025empirical})
            & 2 \bigstrut\\\hline

        Resource label mismatch
            & Hardcoded, outdated label prevents discovery of  intended DB/service.
            & Connection refused; startup failure; unavailability.
            (Reddit 2023 outage~\cite{reddit_pi_day_outage}; report~\cite{hotos_19,cbs_database,k8saf})
            & 2 \bigstrut\\\hline

        Constant misconfiguration
            & Wrong env/DB constant (e.g., array length) distorts control flow.
            & Out-of-bounds access; crashes; error bursts. (CrowdStrike 2024 outage~\cite{crowdstrike2024rca}; report~\cite{ghosh_2022_socc})
            & 2 \bigstrut\\\hline

        Feature-flag ConfigMaps misconfiguration
            & Incorrect shared Flagd rules/config; effects depend on code paths and flag usage.
            & Crashes; pod terminations; CPU/memory pressure. (Spotify 2025 incident~\cite{spotify2025outage}; report~\cite{hotos_19,ghosh_2022_socc})
            & 9 \bigstrut\\\hline

        Deployment manifest error
            & Invalid image tag or replicas{=}0.
            & ImagePullBackOff; service unavailable. (Qury\.io 2020 outage~\cite{dettelback2020quay}; report~\cite{ghosh_2022_socc})
            & 2 \bigstrut\\\hline

        Resource- \& Infrastructure-related fault
            & Chaos Mesh: CPU stress, node-level network outage, JVM corruption.
            & Saturation; packet loss; node/pod failures. (Multiple cases in~\cite{k8saf})
            & 3 \bigstrut\\
        \hlineB{2}
    \end{tabularx}
    }
    \end{threeparttable}
\end{table}

\subsection{Baselines}
We benchmarked \agent against ReAct-style~\cite{yao2023react} RCA agents~\cite{jha2025itbench}:
\begin{itemize}
  \item[\textbf{SRE-Agent (baseline).}] ReAct-style SRE-Agent~\cite{jha2025itbench} that uses standard SRE tools to access the Kubernetes API, traces, metrics, events, and microservices application logs. 
  \item[\textbf{SRE-Agent+CT (baseline+).}] Extends SRE-Agent with \emph{code tools (CT)} for code retrieval, inspection, and summarization, adding code context to the baseline.
  \item[\textbf{\agent (ours).}] The proposed method, \agent, employs LLM-driven reasoning and traversal over program dependence and service dependency graphs.
\end{itemize}

\subsection{Evaluation Metrics}
We evaluated RCA effectiveness with respect to the below metrics following the same definitions as~\cite{jha2025itbench}:
\begin{itemize}
    \item[\textbf{Root cause identification (RCI) Pass@1} (\%):] The percentage of instances for which the identified faulty microservice(s) exactly match the ground truth.
    \REV{In other words, RCI measures whether the agent correctly localizes the root cause to the responsible microservice(s) or associated resource(s).}
  
  \item[\textbf{Root cause reasoning (RCR) Pass@1} (\%):] Pass@1 accuracy of the agent’s RCA reasoning, measured by whether its explanation correctly identifies the faulty statement(s), function(s), or configuration(s) relative to the ground truth, following \textit{Pass@K}~\cite{chen2021evaluating} with \(K=1\).
  \REV{In other words, RCR measures whether the agent pinpoints the underlying fault (e.g., a faulty code-site or configuration) and correctly explains (reasons about) how it led to the incident and the observed symptoms.}
\end{itemize}
We also report RCA efficiency via \textbf{MTTD} (mean time to diagnosis) and \textbf{ATC} (average token consumption).
\REV{See~\tion{overhead} for a detailed explanation of MTTD and ATC.}

\subsection{\agent Implementation}
\agent is implemented in Python 3.12 using the LangGraph agentic framework~\cite{langgraph_2025}.
\agent supports \graph construction for 
microservices written in Python and Java, a scope we chose based on current toolchain availability and the popularity of these languages~\cite{tiobe2025index}.
For evaluation purposes, all implemented faults reside in, or are associated with, codebases available for analysis and supported by \agent.
We will submit our artifacts for evaluation and open-source the implementation.

\section{Evaluation Results}
\label{sec:result}
\begin{table}[t]
    \centering
    \caption{\label{tab:baseline} Evaluation result of baseline and \agent.}
    \resizebox{1.0\linewidth}{!}{
        \begin{threeparttable}
            \begin{tabular}{c|c|c}
                \hlineB{2}
                \rowcolor[HTML]{EFEFEF}
                LLM model & \textbf{RCR Pass@1\%}$\uparrow$ & \textbf{RCI Pass@1\%}$\uparrow$ \bigstrut\\
                \hlineB{2}
                \rowcolor{white}
                \multicolumn{3}{c}{\textbf{SRE-Agent}}                                                     \\\hlineB{2}
                gpt-oss-120b              & 0.00 $\pm$ 0.00                           & 1.53 $\pm$ 1.07              \\
                deepseek-r1               & 0.00 $\pm$ 0.0                            & 9.79 $\pm$ 2.49              \\
                o4-mini                   & 3.42 $\pm$ 1.50                           & 11.64 $\pm$ 2.65             \\
                mistral-medium-3.1        & 9.65 $\pm$ 2.77                           & 21.93 $\pm$ 3.88             \\
                gpt-5-codex               & 0.00 $\pm$ 0.00                           & 11.43 $\pm$ 3.80             \\
                \hlineB{2}
                \rowcolor{white}
                \multicolumn{3}{c}{\textbf{SRE-Agent w/ Code Tools (CT)}}\\\hlineB{2}
                gpt-oss-120b             & 0.68 $\pm$ 0.69       & 1.38 $\pm$ 0.97          \\ 
                deepseek-r1   & 0.00 $\pm$ 0.0       & 11.41 $\pm$ 2.60       \\    
                o4-mini   & 2.68 $\pm$ 1.32       & 14.77 $\pm$ 2.91       \\ 
                mistral-medium-3.1        & 10.07 $\pm$ 2.55    &   20.86  $\pm$ 3.45    \\
                gpt-5-codex    & 0.00 $\pm$ 0.00       & 7.96 $\pm$ 2.54     \\ 
                \hlineB{2}
                \rowcolor{magenta!12}
                \multicolumn{3}{c}{\textbf{\agent (Ours)}}                                                     \\\hlineB{2}
                gpt-oss-120b              & 48.28 $\pm$ 4.15                          & \cellcolor{blue!10}\underline{71.03} $\pm$ 3.77 \\
                deepseek-r1               & 37.50 $\pm$ 4.74                          & 57.69 $\pm$ 4.84             \\
                o4-mini                   & \cellcolor{blue!10}\underline{54.16} $\pm$ 4.15              & 70.14 $\pm$ 3.81             \\
                mistral-medium-3.1        & 4.37 $\pm$ 1.74                           & 23.36 $\pm$ 3.61             \\
                gpt-5-codex               & \cellcolor{blue!20}\textbf{61.54} $\pm$ 4.27                 & \cellcolor{blue!20}\textbf{73.85} $\pm$ 3.85    \\
                \hlineB{2}
            \end{tabular}
            \begin{tablenotes}[para]
                \footnotesize
                Metrics: (1) \textbf{RCR Pass@1}: Pass@1 \% score for root cause reasoning. (2) \textbf{RCI Pass@1}: root cause identification accuracy. Best-performing agent+LLM model is shown in \textbf{bold}, and the second-best is \underline{underlined}. Higher is better.
            \end{tablenotes}
        \end{threeparttable}
    }
\end{table}

We answer the following research questions (RQs):
\bi
\item[\textbf{RQ-1 (\tion{baseline}):}] How accurate is \agent compared to baseline agents for cloud incident RCA?

\item[\textbf{RQ-2 (\tion{ablation}):}] How does incident-centric program context via \graph reasoning contribute to \agent's improvement?

\item[\textbf{RQ-3 (\tion{overhead}):}] What are the diagnosis overheads of incorporating program context in cloud RCA?

\ei
\subsection{Baseline Comparison (RQ1)}
\label{sec:baseline}
For RQ-1, we evaluated \agent by comparing it against the current state of the art, \ie, the SRE-Agent~\cite{jha2025itbench}, 
a ReAct-style RCA agent with default tools, and an extended SRE-Agent with code retrieval and summarization tools (which we call SRE-Agent+CT) to study whether code context can improve on the baseline performance.

\textbf{Evaluation setup.}~We evaluated the agent across five different LLMs: o4-mini, gpt-5-codex, mistral-medium-3.1, deepseek-r1, and gpt-oss-120b. For statistical robustness, we repeated each $\langle\text{LLM},\ \text{scenario}\rangle$ experiment five times with distinct random seeds, totaling 2,250 trajectories.

\textbf{Overall results.} As shown in~\tab{baseline}, \agent consistently performed better than SRE-Agent and SRE-Agent+CT across all LLMs that were evaluated, both on root-cause reasoning (RCR) accuracy and root-cause identification (RCI) accuracy. 
\agent's RCA reasoning accuracy correlates strongly with root-cause identification accuracy, reflecting how human SREs rely on accurate fault pinpointing for effective root-cause analysis.
Reasoning models (gpt-5-codex, o4-mini) exhibited better RCA accuracies overall. Moreover, the strong performance of gpt-5-codex likely stemmed from its superior code-understanding capabilities (demonstrated in~\cite{jimenez2024swebench}), enabling it to better correlate the code context and program behavior with the incident.
The low accuracy of mistral-medium-3.1 was due to its failure to select the correct initial candidates and its inability to follow output format instructions, which resulted in parsing errors.

\textbf{Why did baselines underperform?} 
SRE-Agent underperformed because it lacks code visibility, which is essential for diagnosing code-related scenarios. 
Adding code tools (SRE-Agent+CT) yielded only marginal gains, not because code is unhelpful, but because the search policy remained unguided. Even with code and SRE tools, gpt-5-codex and similar models exhibited \emph{premature closure}: once local evidence seemed to \textit{explain}  alerts (e.g., an \textit{HTTP~500} handler or nearby error traces), they terminated the investigation. 
Without explicit awareness or external imposition of the microservice or program dependencies, these baselines had neither a mechanism for traversing, nor an obligation to traverse the multi-hop dependency chain (which in our scenarios often involved 7–10 hops across different microservices and their hammock blocks) in which the true faults either resided or were indicated.
Notably, despite explicit instructions to invoke code tools in the prompt, SRE-Agent+CT invoked code tools for only 46\% of runs in which code analysis was required for RCA, resulting in low RCA accuracy and inconsistent execution trajectories and highlighting SRE-Agent+CT's fragility in agent tool-calling~\cite{winston_tool-augmentedLLMs_2025,xiong-etal-2025-butterfly}.

In contrast, \agent treats service and code-level dependencies as first-class as it constrains the LLM's reasoning onto the dependency paths in PDG and systematically expands each graph node's neighborhood with admissible graph traversal actions, enforcing iterative context gathering and reasoning reconciliation. 
This graph-aware reasoning prevents early stopping on symptomatic context and steers the investigation to the upstream root cause in code, achieving in-depth RCA. 

\subsection{Ablation Study (RQ2)}
\label{sec:ablation}
\begin{table}[t!]
    \centering
    \caption{\label{tab:ablation} Ablation study results on all scenario instances.}
    \resizebox{1.0\linewidth}{!}{
        \begin{threeparttable}
            \begin{tabular}{c|c|c}
                \hlineB{2}
                \rowcolor[HTML]{EFEFEF}
                \textbf{Agent Variant} & \multicolumn{2}{c}{\textbf{Overall RCA Accuracy}}                             \\  & \textbf{RCR Pass@1\%}$\uparrow$ & \textbf{RCI Pass@1\%}$\uparrow$ \\
                \hlineB{2}
                \agent (Obs. Ctx.)          & 12.93 $\pm$ 2.77                          & 41.50 $\pm$ 4.06          \\
                \agent (Raw Code)      & 32.65 $\pm$ 3.87                          & 59.18 $\pm$ 4.05          \\
                \cellcolor{magenta!12}
                \textbf{\agent (Ours)} & \cellcolor{blue!20}\textbf{61.54} $\pm$ 4.27                 & \cellcolor{blue!20}\textbf{73.85} $\pm$ 3.85 \\
                \hline
                \rowcolor[HTML]{EFEFEF}
                \textbf{} & \multicolumn{2}{c}{\textbf{Data Schema Mismatch}}                             \\ \hline
                \agent (Obs. Ctx.)          &            40.0 $\pm$ 14.14   & \cellcolor{blue!20}\textbf{100.0} $\pm$ 0.0        \\
                \agent (Raw Code)      & 25.0 $\pm$ 18.97                          & 85.0 $\pm$ 14.14          \\
                \cellcolor{magenta!12}
                \textbf{\agent (Ours)} & \cellcolor{blue!20}\textbf{70.0} $\pm$ 8.94                 & 95.0 $\pm$ 8.93 \\
                \hline
                                \rowcolor[HTML]{EFEFEF}
                \textbf{} & \multicolumn{2}{c}{\textbf{Improper Ext. Failure Handling}}                             \\ 
                \hline
                \agent (Obs. Ctx.)          & 10.0 $\pm$ 11.54  & 33.3 $\pm$ 1.6                                  \\
                \agent (Raw Code)      & 36.67 $\pm$ 20.0                          & 43.33 $\pm$ 19.32          \\
                \cellcolor{magenta!12}
                \textbf{\agent (Ours)} & \cellcolor{blue!20}\textbf{85.83} $\pm$ 12.57                 & \cellcolor{blue!20}\textbf{86.0} $\pm$ 12.60 \\
                \hline 
                \rowcolor[HTML]{EFEFEF}
                \textbf{} & \multicolumn{2}{c}{\textbf{Internal Logic Bug}}                             \\ 
                \hline
                \agent (Obs. Ctx.)          & 20.0 $\pm$ 15.49 & 100.0 $\pm$ 0.0                                  \\
                \agent (Raw Code)      & 50.0 $\pm$ 17.89                          & 90.0 $\pm$ 12.65          \\
                \cellcolor{magenta!12}
                \textbf{\agent (Ours)} & \cellcolor{blue!20}\textbf{80.0} $\pm$ 17.89                & \cellcolor{blue!20}\textbf{100.0} $\pm$ 0.0 \\
                \hline 
                                \rowcolor[HTML]{EFEFEF}
                \textbf{} & \multicolumn{2}{c}{\textbf{Resource Label Mismatch}}                             \\ 
                \hline
                \agent (Obs. Ctx.)          & 30.0 $\pm$ 15.4 & 100.0 $\pm$ 0.0                                  \\
                \agent (Raw Code)      & 40.0 $\pm$ 21.91                          & 100.0 $\pm$ 0.00          \\
                \cellcolor{magenta!12}
                \textbf{\agent (Ours)} & \cellcolor{blue!20}\textbf{80.0} $\pm$ 15.5                 & \cellcolor{blue!20}\textbf{100.0} $\pm$ 0.0 \\
                \hline 
                                \rowcolor[HTML]{EFEFEF}
                \textbf{} & \multicolumn{2}{c}{\textbf{Constant Misconfiguration}}                             \\ 
                \hline
                \agent (Obs. Ctx.)          & 0.0 $\pm$ 0.0 & 80.0 $\pm$ 15.0                                  \\
                \agent (Raw Code)      & 20.0 $\pm$ 17.89                          & 90.0 $\pm$ 12.65          \\
                \cellcolor{magenta!12}
                \textbf{\agent (Ours)} & \cellcolor{blue!20}\textbf{90.0} $\pm$ 13.0                 & \cellcolor{blue!20}\textbf{100.0} $\pm$ 0.0 \\
                \hline 
                                \rowcolor[HTML]{EFEFEF}
                \textbf{} & \multicolumn{2}{c}{\textbf{Feature-flag ConfigMaps Misconf.}}                             \\ 
                \hline
                \agent (Obs. Ctx.)          & 0.0 $\pm$ 0.0 & 0.0 $\pm$ 0.0                                  \\
                \agent (Raw Code)      & 45.56 $\pm$ 19.71                          & 54.44 $\pm$ 19.71          \\
                \cellcolor{magenta!12}
                \textbf{\agent (Ours)} & \cellcolor{blue!20}\textbf{62.96} $\pm$ 16.67                 & \cellcolor{blue!20}\textbf{65.74} $\pm$ 18.16 \\
                \hlineB{1} 
            \end{tabular}
            \begin{tablenotes}[para]
                \footnotesize
                We used the same RCA accuracy metrics as  in~\Cref{tab:baseline}.
            \end{tablenotes}
        \end{threeparttable}
    }
\end{table}
For this RQ, we conducted an ablation study to assess the impact of \agent's core components on RCA effectiveness by evaluating the following \agent variants.
\bi
\item[\textbf{\agent (Obs. Ctx.):}] The observability context-only variant disables \graph-traversal and program context ($C_P$), so only the incident context and observability context ($c_i$) are used in the RCA decision in~\tion{entity_role}.
This experiment characterized the performance gains resulting from the use of program context. 

\item[\textbf{\agent (Raw Code):}] The Raw Code variant skips \graph construction (\tion{pdg}) and LLM-driven \graph traversal (\tion{traver}) when constructing $C_P$ and instead uses an LLM with a modified $\psi_{syn}$ prompt (\tion{traver}) to consolidate and generate program context directly from aggregated raw code files. We limited the raw code to 800k characters ($\sim$200k tokens) to avoid overflowing the LLM's context window.
\ei

\textbf{Evaluation setup.}~We evaluated these \agent variants using the best-performing LLM for \agent--gpt-5-codex. For statistical robustness, we repeated each $\langle\text{variant},\ \text{scenario}\rangle$ experiment five times with distinct random seeds, yielding a total of 300 trajectories.

\begin{figure}[t!]
    \centering
    \includegraphics[width=1.0\linewidth]{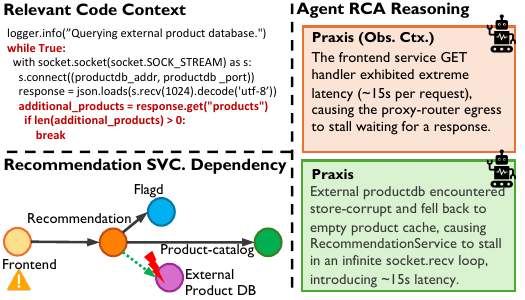}
    \caption{\label{fig:praxis_ablation_nocode} RCA reasoning of \agent (Obs. Ctx.) and \agent. Nodes are microservices; solid arrows are dependencies; the green dotted arrow is a dependency with missing traces and had to be derived from program context. SVC. = Service.}
\end{figure}

\textbf{Ablation results.}
Table~\ref{tab:ablation} shows that augmenting the observability context \(c_i\) with program context \(C_P\) (PDG construction + LLM-guided traversal) is the primary driver of RCA effectiveness. Relative to \agent (Raw Code), \agent improves overall RCR Pass@1 by 28.8 percentage points, from 32.7\% to 61.5\% (an 88.5\% improvement), and RCI Pass@1 by 14.7 percentage points, from 59.2\% to 73.9\% (a 24.8\% improvement). Against \agent (Obs.\ Ctx.), the gains are larger, at 48.6 percentage points and 32.4 percentage points for RCR Pass@1 and RCI Pass@1, respectively. 

\textbf{Why do program context and LLM-driven \graph help?} 
Similar to the SRE-Agent baseline, \agent without program context (\agent (Obs. Ctx.)) failed to identify root causes within, or hinted by, the internal code logic, resulting in superficial symptom-level RCA. 
Although the \agent variant without \graph (\agent (Raw Code)) retains service-level dependency awareness, the absence of \graph-guided diagnosis halved its accuracy compared to that of the full approach. 
Lacking explicit awareness of program-level dependencies, \agent (Raw Code) cannot effectively focus its reasoning on code regions relevant to the incident. Instead, it ``fixated'' on ostensibly faulty regions---code that appears erroneous on the surface (e.g., manually thrown exceptions)---that are not pertinent to the observed failure. 
With code context exceeding 200k tokens at times, \agent (Raw Code) exhibited \textit{needle in a haystack}~\cite{gola_multi_needle_2024} and \textit{context rot}~\cite{hong2025contextrot} phenomena, leading to degraded RCA accuracy.
These findings confirm that \graph-guided diagnosis is essential: it enforces the alignment of observability data with concise, accurate code context structurally presented to the LLM, grounding the RCA in relevant code paths and code regions to yield the observed performance gains.

\textbf{Illustration.} 
We illustrate the RCA reasoning specificity improvement using an ``improper handling of external dependency failure'' scenario. 
Here, a degraded external database returned empty responses, triggering a silent retry loop in the Recommendation service that manifested solely as high latency, without explicit error logs. 
As shown in \Cref{fig:praxis_ablation_nocode}, \agent without program context failed to detect this behavioral pattern, wrongly identifying the high latency, a symptom, as the root cause. 
Conversely, utilization of program context enabled \agent to identify the retry logic and uncover the database dependency, correctly isolating the underlying storage issue (storage corruption $\rightarrow$ empty product list) as the true root cause.

\textbf{Per-scenario trends.} \agent's improvements hold across most scenario types. For example, for \textit{Constant Misconfiguration}, \agent attained \(\mathbf{90\%}\) RCR and \(\mathbf{100\%}\) RCI while other agents achieved around 0\%--20\% RCR. 
A minor regression appeared in \textit{Data Schema Mismatch} (RCI Pass@1 100.0\% $\rightarrow$ 95.0\%), likely because error signals in logs occasionally allow the observability-only variant (\agent (Obs. Ctx.)) to correctly identify the faulty microservice without code analysis, although it is unable to arrive at the correct root cause. 
\textit{Feature-flag ConfigMaps misconfiguration} scenarios remain challenging (RCI \(\approx 66\%\)), likely due to insufficient observability data for \agent to exploit during the \graph-traversal and RCA decision-making.

\subsection{Diagnosis Overheads (RQ3)}
\label{sec:overhead}
For this RQ, we evaluated the RCA overheads of SRE-Agent baselines, \agent, \agent (Obs. Ctx.), and \agent (Raw Code) 
using mean time to completion (MTTC) in seconds and average token consumption (ATC) in thousands of tokens, 
measured across all scenarios using the best-performing LLMs.

\textbf{Evaluation setup.} To account for differences in RCA reasoning accuracy, we normalized both MTTC and ATC by RCR Pass@1\% to assess diagnosis efficiency:
$\text{\textit{Normalized metric}} = \frac{\text{\textit{MTTC or ATC}}}{\text{RCR Pass@1\%}}$.
The resulting normalized MTTC (i.e., mean time to diagnosis or MTTD) and normalized ATC (Effective ATC) reflect time and token costs per successful diagnosis.
\REV{Here, MTTC is the average time to complete a diagnosis, regardless of correctness. MTTD is the normalized MTTC and represents the expected time to reach a \emph{correct} diagnosis. 
Thus, MTTD jointly captures duration and accuracy: an agent that completes diagnosis faster but is less accurate may still require more time, in expectation, to reach a correct diagnosis, resulting in an MTTD longer than that of a more accurate agent.
The same interpretation applies to ATC and Effective ATC.}

\Cref{tab:overheads} shows that although SRE-Agent has lower raw MTTC and ATC than \agent, their MTTDs, which include LLM backend response times, are comparable.
However, ReAct consumed $5.1\times$ more tokens per successful diagnosis, indicating much lower token efficiency.
When we compare \agent with its ablated variants (\agent (Raw Code)) and \agent (Obs. Ctx.)), construction of program context via LLM-driven \graph traversal reduced MTTD and effective ATC by up to $85.9\%$ and $83.5\%$, respectively, highlighting their significant contributions to efficient RCA.
Notably, LLM-driven \graph traversal reduced ATC by $6.1\times$ compared to providing raw code as plain text in the prompt, as in~\cite{rcacp_chen2024,li2025coca_icse}.
\agent's MTTD of 1,474 seconds is substantially shorter than that of manual RCA, which can take several hours for incidents with similar root causes.

\begin{table}[t]
    \centering
    \caption{\label{tab:overheads} Diagnosis overheads across all scenario instances.}
    \resizebox{1.0\linewidth}{!}{
        \begin{threeparttable}
            \begin{tabular}{c|c|c|c|c}
                \hlineB{2}
                & \textbf{MTTC}$\downarrow$ & \textbf{ATC}$\downarrow$ & \textbf{MTTD}$\downarrow$ & \textbf{Eff. ATC}$\downarrow$  \\
                \hlineB{2}
                SRE-Agent       & \cellcolor{blue!10}\underline{99.69}                             & {85.39k} & \cellcolor{blue!10}\underline{1,033.06}    & 884.87k             \\ 
                SRE-Agent+CT       & \cellcolor{blue!20}\textbf{90.18}                             & \cellcolor{blue!10}\underline{85.03k} & \cellcolor{blue!20}\textbf{842.80}    & 794.67k             \\ \hlineB{2}
                \agent (Obx. Ctx.)          & 1,321.02                                    & \cellcolor{blue!20}\textbf{75.59k}    & 10,475.97            & \cellcolor{blue!10}\underline{599.44k} \\
                \agent (Raw Code)      & {838.44}                                     & 329.53k            & 2,567.96             & 1,009.28k            \\
                \cellcolor{magenta!12}
                \textbf{\agent (Ours)} & 907.43                                     & 102.44k            & {1,474.54} & \cellcolor{blue!20}\textbf{166.46k}    \\
                \hlineB{2}
            \end{tabular}
            \begin{tablenotes}[para]
                \footnotesize
                Metrics: (1) \textbf{MTTC} = mean time to completion in seconds; (2) \textbf{ATC} = average token consumptions in tokens; (3) \textbf{MTTD} =  mean time to diagnosis; (4) \textbf{Eff. ATC} =  effective ATC. Best-performing variant is shown in \textbf{bold}; second-best is \underline{underlined}. Lower is better.
            \end{tablenotes}
        \end{threeparttable}
    }
\end{table}

\subsection{Discussions and Future Work}
\label{sec:diss}
\textbf{Scenarios with weak observability.}
\agent's effectiveness is contingent on the availability and quality of observability data to anchor its \graph traversal.
Sparse or ambiguous traces, logs, metrics, or events can limit \agent's ability to accurately align runtime error signals with corresponding dependency paths to traverse. 
When a fault offers no direct or clear observable signal at the application layer to anchor RCA decision-making, \agent tends to falsely attribute the incident to fault-free code paths or configurations, leading to false positives.
While collecting exhaustive observability data could mitigate this, it would incur significant production overheads.
We explicitly identify the fault categories most vulnerable to this limitation in our scope discussion below.

\textbf{Graph representation.} 
\agent's effectiveness relies on the correctness and completeness of the dependency graphs. 
Dependency graphs that are incorrect or incomplete (e.g., an outdated \graph, or a missing microservice in \sgraph) hinder \agent's ability to diagnose a root cause. 
Because \agent diagnoses faults by traversing dependency edges anchored on observability signals, missing edges break the structural path from symptom to root cause: observability may reveal the symptomatic component, but without the corresponding edge, the agent cannot reach the true root-cause node.
For isolated nodes or a complete absence of dependency structures (e.g., infrastructure/resource dependencies), traversal terminates prematurely at the upstream symptom, reducing RCA accuracy.
These implications explain \agent's practical scope and limitations in its present form, described in the next paragraph.

\REV{ %
\textbf{\agent's scope and limitations.}
Our evaluation (\Cref{tab:ablation}) shows that \agent is effective at diagnosing (1) code-related incidents that fall under the fault models \textit{data schema mismatches, logical bugs, failure to handle external dependencies, and resource/address/label mismatches}, and (2) misconfiguration-related incidents caused by \textit{misconfigured variables/constants in code or in ConfigMaps} (see~\tion{ablation},~\Cref{tab:ablation}). 
Collectively, these fault models represent the vast majority of code-related and misconfiguration incidents observed in production~\cite{ghosh_2022_socc,hotos_19,gunawi2014cloudbug}.
\agent works well here as these faults emit observability signals and either are directly visible in code or can be more accurately pinpointed via code examination (e.g., a misconfiguration altering execution paths).

As discussed, \agent's effectiveness depends on the availability and quality of observability signals and the completeness of its dependency graphs.
Weak observability signals (e.g., missing traces or ambiguous logs) and incomplete or incorrect graphs can hinder diagnosis.
This explains why \agent and its variants achieve near-zero accuracy on \textit{deployment-manifest} and \textit{resource/infrastructure}-related incidents: those incidents lack direct application-layer observability and their dependency structures are not captured in \agent's graphs. 
We exclude these incidents from~\Cref{tab:ablation} because all ablated variants behave and perform identically, so including them would not change the ablation comparison.

For the same reason, we consider \textit{concurrency/timing} bugs out of scope: \agent in its current form lacks dynamic runtime analysis to observe microservices' internal runtime state and does not encode timing-dependency structures between microservices.
These limitations are not inherent to the graph-based approach, and \agent can be extended to address them, as discussed in the ``Future work'' section.
Finally, while \agent may generalize to fault types beyond those evaluated, its effectiveness on such faults remains unverified.%
}

\textbf{Access to source code.}
\agent's gains from \graph traversal are naturally limited when microservice code is unavailable for graph construction or code access is substantially limited (e.g., extensive reliance on proprietary third-party libraries), in which case \agent must rely primarily on observability signals for diagnosis.
\REV{%
In practice, partial code access is common (and matches our evaluation): Devs/SREs can access code developed in-house but not proprietary third-party binaries. 
Even with partial code access, \agent can exploit clues in the available code captured by the \graph (e.g., call sites and configuration-controlled branches) to localize the root cause to external dependencies, thereby achieving higher RCA effectiveness than observability-only variants, see~\Cref{fig:praxis_motivation} and~\tion{ablation} for  examples.%
}

\REV{
\textbf{Static code analysis.} 
An inherent limitation of static code analysis is its inability to accurately capture dynamic bindings (e.g., Java reflection), which can leave unlinked gaps in the \graph.
Although \agent cannot eliminate this limitation, its hierarchical hammock-block traversal alleviates it by leveraging higher-level structural relations (e.g., class-level) to potentially bypass missing links at lower levels.
Conversely, conditioning too strongly on static code context risks unintended bias, where \agent may mistakenly attribute faults to code blocks when the incident actually stems from external factors (e.g., deployment or resources). 
This risk can be mitigated by incorporating runtime instrumentation to confirm whether implicated code paths were indeed executed at the time of the incident.
}

\REV{%
\textbf{\graph construction overheads and continuous deployment.}
Regenerating all \graphs in our case took 116.42 seconds.
\agent builds \graphs by using a syntactic parser (e.g., Tree-sitter) to extract hammock blocks, and running CLDK-provided static analysis (e.g., WALA/RTA for Java and CodeQL/Scalpel for Python)~\cite{rahul2024cldk}.
In practice, construction time is dominated by the static-analysis backend, and lightweight backends (e.g., the RTA~\cite{bacon1996rta} backend used by CLDK) scale near-linearly with microservice program size~\cite{grove_2001_rta}.
Subsequent \graph construction takes $O(N+E)$, where $N$ is the number of hammock blocks and $E$ is the number of dependency edges extracted.
Going forward, \agent will be integrated with a CI/CD\footnote{\textit{CI/CD} stands for \textit{continuous integration and continuous deployment}\cite{redhat_cicd_2025}.} pipeline. 
\agent can avoid fully regenerating the \graphs because (1) \graphs can be constructed and parallelized per microservice, and (2) each PDG update can be  incremental~\cite{hammer2015increment,souter2001incre}, recomputing only nodes and transitive dependence regions affected by code changes. Moreover, caching versioned PDGs by commit SHA enables fast reuse and comparison.%
}

\REV{%
\textbf{Extensibility to other languages and platforms.} \agent is extensible beyond Python/Java and Kubernetes. \sgraph construction requires discovery of the runtime inter-service topology and dependencies, capabilities already supported by observability stacks (e.g., OpenTelemetry, Datadog) across platforms (e.g., OpenShift, ECS).
\graph generation requires a syntactic parser for extracting hammock blocks and a static-analysis backend to derive control/data-flow edges; thus, specific language support is primarily limited by the availability of such toolchains.
Because \agent constructs \graph using Tree-sitter~\cite{tree_sitter_zenodo_2025} and CLDK~\cite{rahul2024cldk}, which provide broad multi-language support, it can be readily extended to additional languages, including Go, Rust, TypeScript, and C/C++. %
}

\textbf{Future work.} 
Future iterations of \agent will be extended to diagnose resource-, infrastructure-, and deployment-related incidents by augmenting the current dependency graphs with resource/infrastructure and deployment-manifest graphs that explicitly capture those dependencies.
\REV{Furthermore, \agent will incorporate dynamic program analysis to address static-analysis gaps and enable the diagnosis of concurrency bugs. By leveraging lightweight, strategic runtime instrumentation (e.g., at database calls and gRPC handlers) alongside dynamic tracing with sampling~\cite{Liu2017DCatch,lu2018cloudraid,yuan2020effcon}, \agent can capture dynamic bindings and gain crucial visibility into internal microservice states. This runtime context will help reduce false positives for resource, infrastructure, and deployment faults, while simultaneously bringing concurrency bugs into scope.}
Ultimately, these techniques will ground the static graph traversal in verifiable runtime behavior, enabling more comprehensive incident coverage and precise RCA.

\section{Related Work}
\textbf{Agentic approach for cloud incident RCA.}
Recent agentic RCA approaches~\cite{roy2024exploring_rca_agents,jha2025itbench,xu2025openrca,chen2025aiopslab,chen2025stratus} adopted the ReAct paradigm~\cite{yao2023react} to analyze observability data but lack structured reasoning workflow and program context, limiting their effectiveness on software and misconfiguration faults.
Prior approaches that incorporate program or code context~\cite{wang2024rcagent,li2025coca_icse} treat code as plain text in prompts rather than explicitly exploiting its inherent structure, as they do not perform \graph-guided reasoning.
\agent advances beyond those efforts by performing LLM-driven reasoning over service dependency and program dependence graphs, unifying incident, observability, cloud, and program context for scalable, end-to-end RCA.

\textbf{AI/ML for specific RCA workflow automation.} 
AI/ML techniques have long supported incident RCA through specialized models for anomaly detection, fault localization, and diagnosis~\cite{deeplog2017,anomalydetectionMS2019,zeyan2021,2021_sage,jha2020_kai,cloudRCA2021}.
Recent work has employed LLMs for incident querying, understanding, and classification~\cite{xpert2024,xlifecycle_2024,llmoutageunder2023,zhang2024flash,rcacp_chen2024,faSmodel2024,FoA2025}, typically leveraging prior knowledge (e.g., TSGs, SOPs) and requiring SRE supervision.
\agent performs automated RCA end-to-end, and those methods can be considered complementary to it.

\textbf{LLM for graph reasoning and traversal.} 
LLM-driven graph reasoning/traversal is an emerging field~\cite{bei2025graphsmeetaiagents} that can be applied to knowledge retrieval, question-answering~\cite{jin2024graphcot,sun2024thinkongraph,pog_2025,knowledgegraph2025}, code search and localization~\cite{chen-etal-2025-locagent}, and execution path reconstruction~\cite{pu2025errorprism}. Though graph models have been established in systems/AIOps~\cite{graphRCA2024}, LLM-driven graph reasoning in this domain is still underexplored, and we have demonstrated its potential through \agent.
\section{Conclusion}
This paper proposes \agent, an agentic approach that reasons over microservice and program dependency graphs for a comprehensive, in-depth root-cause analysis. Our evaluation of \agent on 30 scenarios that span software, configuration, deployment, and resource failures demonstrates that \agent achieves up to $6.3\times$ better root-cause reasoning  accuracy, $3.4\times$ higher entity identification accuracy, and a $5.3\times$ reduction in token consumption compared to state-of-the-art agentic baselines.

\section*{Acknowledgment}
We thank the reviewers, and R. Arora, Bhavya, N. Zheutlin, P. T. Isaza, J. Ahn, A. Paradkar, L. Shwartz, R. K. Kottapalli, A. D. Angelis, A. Patke, P. Cao, Z. Zheng, and J. Applequist for technical input and feedback, and H. C. Fairow, S. Weick, K. Atchley, Y. Deng, R. Pavuluri, M. Vukovic, X. Liu, D. Sow, and N. Fuller for administrative support. This work is supported by the IBM-Illinois Discovery Accelerator Institute (IIDAI) and NSF grant 2530738. Any opinions expressed here are those of the authors and do not necessarily reflect the views of IBM or NSF.

    \bibliographystyle{IEEEtran}
    \bibliography{bibliography.bib}

\end{document}